\documentstyle[11pt,epsfig]{article}

\begin{document}
\large

\def\lsim{\mathrel{\rlap{\lower3pt\hbox{\hskip0pt$\sim$}}
    \raise1pt\hbox{$<$}}}         %less than or approx. symbol
\def\gsim{\mathrel{\rlap{\lower4pt\hbox{\hskip1pt$\sim$}}
    \raise1pt\hbox{$>$}}}         %greater than or approx. symbol
\def\dblint{\mathop{\rlap{\hbox{$\displaystyle\!\int\!\!\!\!\!\int$}}
    \hbox{$\bigcirc$}}}
\def\ut#1{$\underline{\smash{\vphantom{y}\hbox{#1}}}$}

\newcommand{\beq}{\begin{equation}}
\newcommand{\eeq}{\end{equation}}
\newcommand{\dem}{\Delta M_{\mbox{B-M}}}
\newcommand{\dega}{\Delta \Gamma_{\mbox{B-M}}}

\newcommand{\ind}[1]{_{\begin{small}\mbox{#1}\end{small}}}
\newcommand{\WA}{{\em WA}}
\newcommand{\SM}{Standard Model }
\newcommand{\QCD}{{\em QCD }}
\newcommand{\KM}{{\em KM }}
\newcommand{\hscale}{\mu\ind{hadr}}
\newcommand{\sG}{i\sigma G}

\newcommand{\MS}{\overline{\mbox{MS}}}
\newcommand{\pole}{\mbox{pole}}
\newcommand{\aver}[1]{\langle #1\rangle}

\newcommand{\appa}{\mbox{\ae}}
\newcommand{\CP}{{\em CP } }
\newcommand{\fy}{\varphi}
\newcommand{\hi}{\chi}
\newcommand{\al}{\alpha}
\newcommand{\as}{\alpha_s}
\newcommand{\gf}{\gamma_5}
\newcommand{\gm}{\gamma_\mu}
\newcommand{\gn}{\gamma_\nu}
\newcommand{\be}{\beta}
\newcommand{\ga}{\gamma}
\newcommand{\de}{\delta}
\renewcommand{\Im}{\mbox{Im}\,}
\renewcommand{\Re}{\mbox{Re}\,}
\newcommand{\GeV}{\,\mbox{GeV}}
\newcommand{\MeV}{\,\mbox{MeV}}
\newcommand{\matel}[3]{\langle #1|#2|#3\rangle}
\newcommand{\state}[1]{|#1\rangle}
\newcommand{\ra}{\rightarrow}
\newcommand{\ve}[1]{\vec{\bf #1}}

\newcommand{\rhs}{{\em rhs}}
\newcommand{\pp}{\langle \ve{p}^2 \rangle}

\newcommand{\BR}{\,\mbox{BR}}
\newcommand{\La}{\overline{\Lambda}}

\vspace*{.7cm}
\begin{flushright}
\large{
CERN-TH.7207/94\\
UND-HEP-94-BIG\hspace*{0.1em}04}\\
%\today \\
\end{flushright}
\vspace{1.2cm}
\begin{center} \LARGE {\bf CP VIOLATION IN
BEAUTY DECAYS -- \\
THE STANDARD MODEL PARADIGM\\
OF LARGE EFFECTS
\footnote{Invited lecture given at the VIII
Rencontres de Physique de la Vallee d'Aoste, La Thuile,
March 1994}}
\end{center}
\vspace*{.4cm}
\begin{center} \Large
I.I. Bigi\\
\vspace*{.4cm}
{\normalsize {\it Theoretical Physics Division, CERN, CH-1211
Geneva 23, Switzerland \footnote{During the academic year
1993/94}}\\
and\\
{\it Dept. of Physics,
University of Notre Dame du
Lac, Notre Dame, IN 46556 \footnote{Permanent address} } \\
{\it e-mail address: IBIGI@CERNVM, BIGI@UNDHEP}}
\vspace{.4cm}
\end{center}
\thispagestyle{empty} \vspace{.4cm}

\centerline{\Large\bf Abstract}
\vspace{.4cm}
The Standard Model contains a natural source for CP asymmetries
in weak decays, which is described by the
KM mechanism. Beyond $\epsilon _K$
it generates only elusive manifestations of CP violation in
{\em light-}quark systems. On the other hand
it naturally leads to large
asymmetries in certain non-leptonic beauty decays. In particular
when $B^0-\bar B^0$ oscillations are involved,
theoretical uncertainties in the
hadronic matrix elements either drop out or can be
controlled, and one predicts asymmetries well in
excess of 10\% with high parametric reliability. It is
briefly described how the KM triangle can be determined
experimentally and then subjected to sensitive consistency
tests. Any failure would constitute indirect, but
unequivocal evidence for the intervention of New Physics;
some examples are sketched. Any
outcome of a comprehensive program of CP studies in $B$ decays
-- short of technical failure -- will provide us with fundamental
and unique insights into nature's design.

\vspace{3cm}
\noindent CERN-TH.7207/94 \\
\noindent March 1994
\newpage
\large
\addtocounter{footnote}{-3}
\addtocounter{page}{-1}
\section{Introduction}

Exact CP invariance holds neither in nature
-- the decay
$K_L\ra \pi \pi$ has been observed -- nor
in the Standard Model (hereafter
referred to as SM): with three
quark families there arises
one irreducible weak phase $\delta _{KM}$.
The KM ansatz has scored some at least
qualitative successes in that it naturally
predicts
a small value for $\epsilon _K$,
a tiny one for $\epsilon '$ and even tinier
ones for the electric dipole moments of neutrons
and electrons. As CP violation is ascribed to
the coherent interplay between the three quark
families, it comes as no surprise that
CP asymmetries are expected
to be larger in beauty than in strange
decays since the three families can affect the
former transitions more directly than the latter.
This argument can easily be sharpened and made
more specific -- once a general piece of information
from the data is used. The KM matrix which
{\em a priori} has to be merely unitary, exhibits
a very peculiar
hierarchical and almost symmetrical structure
\cite{WOLF}:
$$V_{KM}\simeq
\left( \begin{array}{ccc}
   1              &\lambda           &\sim \lambda ^3\\
   -\lambda       & 1                &\simeq \lambda ^2\\
   \sim \lambda ^3&\simeq -\lambda ^2&    1
       \end{array}
\right)\; . \; \; \;
\lambda =\sin \theta _c     \eqno(1)$$
The unitarity of the $3\times 3$ matrix $V_{KM}$ leads to
nine independent algebraic relations between the matrix
elements. Three of those are of the form
$$\sum _{i=u,c,t}|V_{ij}|^2=1\; \; \; ,
\; j=d,s,b\; . \eqno(2)$$
They represent weak universality of the charged current
couplings and are thus of fundamental importance -- yet
they tell us nothing (directly) about weak phases. The other
six relations are expressed as follows:
$$\sum _{i=u,c,t}V^*(ij)V(ik)=0\; \; \; ,
\; j\neq k=d,s,b\; ,
\eqno(3)$$
$$\sum _{j=d,s,b}V^*(ij)V(kj)=0\; \; \; ,
\; i\neq k=u,c,t\; .
\eqno(4)$$
Equations (3) and (4)
represent six triangle relations in the complex
plane with their angles constituting observable
weak phases. It can easily be shown that all six triangles possess
the {\em same area} reflecting the fact that
with three quark families there is a single
KM phase $\delta _{KM}$.
If that phase vanishes all
these triangles collapse into straight lines \cite{CECILIA}.

Two of these triangles are highly `squashed', two sides
having length $\sim {\cal O}(\lambda )$ and the third one being
very tiny, namely $\sim {\cal O}(\lambda ^5)$. The
corresponding weak phase is either minute, or its impact on a
transition highly diluted. This describes the situation in
$K$ and in $D$ decays. The next two triangles are still
squashed, although to a lesser degree; i.e. the
lengths of two of its sides are $\sim {\cal O}(\lambda ^2)$
and of the third $\sim {\cal O}(\lambda ^4)$.

The remaining two triangles have all three sides of
${\cal O}(\lambda ^3)$, i.e. of comparable length.
It turns out that
those two triangles are practically the same. Thus there
exists a single triangle, where all three angles are
{\em naturally large}, which is sketched in
Fig. ~\ref{F1}.

\begin{figure}
\begin{center}
\mbox{\epsfig{file=7207fg1.eps,height=5cm}}
%\vspace*{-1cm}
\end{center}
\caption[]{KM triangle with three
naturally large angles}
\label{F1}
\end{figure}

The baseline AC
describes KM-favoured $B$ decays, the
side BC KM-suppressed $B$ decays and the side AB
$B_d-\bar B_d$ oscillations; i.e. the one unitarity
triangle where all three angles are large
is the one most relevant for $B$ transitions! This
general result is
quite stable under variations in the experimental input
numbers: it follows from the general pattern of the
KM matrix sketched above;
changing the KM parameters by a factor of 2 or
so will of course affect the size of the weak phases,
but it will not make them universally small! For this
to happen,
the hierarchical structure of $V_{KM}$, as stated in eq. (1),
had to disappear.

Weak phases can of course lead to an observable CP asymmetry only
if two different amplitudes contribute {\em coherently} to the
same process. This prerequisite can be met in two ways:

(i) $P^0-\bar P^0$ {\em oscillations}:
$B^0-\bar B^0$ oscillations have been found to proceed `speedily',
and on the semi-quantitative level this is well understood
theoretically. `CP violation involving $B^0-\bar B^0$
oscillations' thus provides us with an optimal
laboratory for study.

(ii) {\em Direct CP violation}: This
can arise both in neutral and
in charged $B$ decays (and in $\Lambda _b$ decays as well), and
it can materialize in two ways. ($\alpha $) A difference
between CP conjugate total rates can occur: those rates are
necessarily KM-suppressed, and the intervention of non-trivial
final-state interactions (FSI) is required. Effects due to FSI
are typically beyond our theoretical control; FSI thus
represent a `necessary evil'. ($\beta$) T-odd correlations
among the decay products can arise; e.g. in the decay
$B^-\ra \rho ^-\rho ^0\ra (\pi ^-\pi ^0)(\pi ^+\pi ^-)$
one can search for a non-trivial T-odd correlation among the pion
momenta: $\langle \vec p_-\cdot (\vec p_+ \times \vec p_-)\rangle
\neq 0$. Yet one has to keep in mind that FSI can fake such
an effect, even in the absence of CP violation, as is well known from
analogous studies in $K$ decays: Coulomb forces can generate a
transverse polarization for muons in $K_L\ra \mu ^{\pm}
\pi ^{\mp}\nu$ (but not in $K^{\pm}\ra \mu ^{\pm}\pi ^0\nu$).
FSI are then to be viewed as a `nuisance' here. In the following
I will not discuss T-odd correlations any further; for a
recent discussion of such effects, see e.g. ref.
\cite{TODD}.

In summary: the Standard Model contains a paradigm for
CP violation; it is based on the KM ansatz and it
unequivocally predicts large CP asymmetries in $B$
decays.

\section{CP Phenomenology for $B^0$ Decays}

\subsection{Qualitative Discussion}

Consider a final state $f$, which is common to both $B^0$
and $\bar B^0$ decays; within the SM, $f$ is necessarily
a non-leptonic channel. A state that initially is a
$B^0$ meson can then produce $f$ at time
$t$ \footnote{Throughout this talk `time'
means actually {\it proper} time.}
in two different ways: it can decay into $f$ directly or
it can first oscillate into $\bar B^0$ and then decay. These
two reactions are indistinguishable and thus coherent.
Therefore a different rate can result when starting from
an initial $\bar B^0$ leading to the CP conjugate final
state $\bar f$ at time $t$:
$$\Gamma (B^0\ra f;t)=e^{-\Gamma _Bt}G_f(t)
\; , \eqno(5)$$
$$\Gamma (\bar B^0\ra \bar f;t)=
e^{-\Gamma _Bt}\bar G_{\bar f}(t)\; . \eqno(6)$$
The superposition principle of quantum mechanics allows us
to write down for the functions $G_f(t)$ and
$\bar G_{\bar f}(t)$ \cite{CBS,BSKU}:
$$G_f(t)=|T_f|^2[(1+|\bar \rho _f|^2)+
(1-|\bar \rho _f|^2)\cos (\Delta mt)+
2\sin (\Delta mt)\Im \frac{q}{p}\bar \rho _f]
\eqno(7)$$
$$\bar G_{\bar f}(t)=|\bar T_{\bar f}|^2
[(1+|\rho _{\bar f}|^2)+
(1-|\rho _{\bar f}|^2)\cos (\Delta mt)+
2\sin (\Delta mt)\Im \frac{p}{q}\rho _{\bar f}]
\eqno(8)$$
$$\bar \rho _f=\frac {\bar T(\bar B\ra f)}{T(B\ra f)}\; ,
\; \rho _{\bar f}=
\frac {T(B\ra \bar f)}{\bar T(\bar B\ra \bar f)}\; ,
\; \frac{q}{p}=\sqrt {\frac{M^*_{12}-i\Gamma ^*_{12}/2}
{M_{12}-i\Gamma _{12}/2}}\; , \eqno(9)$$
where $T_f$ and $\bar T_f$ denote the decay amplitudes
of $B$ and $\bar B$ mesons, respectively, into $f$.
Two special cases will help to elucidate
the meaning of these expressions.

($\al$) $\Delta m=0$, i.e. no $B^0-\bar B^0$ oscillations:
$$G_f(t)=2|T_f|^2\; , \bar G_{\bar f}(t)=2|\bar T_{\bar f}|^2
\; \;
\leadsto  \frac{\bar G_{\bar f}(t)}{G_f(t)}=
\frac{|\bar T_{\bar f}|^2}{|T_f|^2}\, ,
\eqno(10)$$
i.e. a possible CP asymmetry -- $\bar G_{\bar f}(t)\neq
G_f(t)$ -- is given completely by a difference in the
decay amplitudes; this `direct CP violation'
is {\it independent} of the time of decay.

($\be$) Let $f$ be a CP eigenstate -- like
$\psi K_S$ -- with $|\bar \rho _f|^2=1$. Then
one finds from eqs. (7) and (8):
$$G_f(t)=2|T_f|^2(1+\sin (\Delta mt)
\Im \frac{q}{p} \bar \rho _f)\eqno(11)$$
$$\bar G_f(t)=2|T_f|^2(1-\sin (\Delta mt)
\Im \frac{q}{p} \bar \rho _f)\eqno(12)$$
$$\frac{G_f(t)}{\bar G_f(t)}-1=
\frac{2\sin (\Delta mt)\Im (q/p)\bar \rho _f}
{1-\sin (\Delta mt)\Im (q/p)\bar \rho _f}
\, , \eqno(13)$$
i.e. CP asymmetries involving $B^0-\bar B^0$ oscillations
exhibit a very peculiar dependance on the time of decay!

These two cases are illustrated in Fig. ~\ref{F2},
where
$(G_f(t)/\bar G_f(t))-1$ has been plotted:
the straight {\it dotted} line represents the case of
{\it no} CP asymmetry,
the straight {\it dashed} line a
{\it direct} CP asymmetry of 20 \%
and the solid line a CP asymmetry involving
$B^0-\bar  B^0$ oscillations with
$\Im (q/p)\bar \rho _f=0.20$.

\begin{figure}
\begin{center}
\mbox{\epsfig{file=7207fg2.eps,height=6cm}}
%\vspace*{-1cm}
\end{center}
\caption[]{CP asymmetry as
a function of the normalized time of decay
$\Delta m_B t$}
\label{F2}
\end{figure}

\vspace{.3cm}

Several observations on CP violation involving
$B^0-\bar B^0$ oscillations should be noted here:

\vspace{.3cm}

(i) When a final state $f$ is common to $B^0$
and $\bar B^0$ decays, it is {\em ipso facto} `flavour-blind'.
Therefore -- analogously to the situation with $K_L\ra
\pi ^+\pi ^-$ decays in the CPLEAR experiment --
independent flavour tagging is required (for an exception to
this general rule see point (viii) below); e.g.:
$$B^+\; \; \bar B^0\ra \psi K_S\; \; \; \; \; vs.
\; \; \; \; \; B^-\; \; B^0\ra \psi K_S$$

(ii) The quite peculiar time dependence of
the CP asymmetry, as illustrated in Fig. 2, will serve as a
valuable guardian against background fakes.

(iii) There exists no urgent need to measure
`quick' $B$ decays with $t<\tau _B$.

(iv) There is, however, the danger that
overly rapid oscillations might wash out CP
asymmetries in $B_s$ decays.

(v) It makes eminent sense to compare the rates
for $B_{neutral}\ra l^-\bar \nu D^{+*}$ and for
$B_{neutral}\ra l^+\nu D^{-*}$: in doing so one probes
for a bias in the detector, since there can be no CP
asymmetry in this transition!

(vi) For the `Kabir Test of T invariance'
one compares the oscillation rate for
$B^0\Rightarrow \bar B^0$ with
that for $\bar B^0\Rightarrow B^0$.
To infer the
flavour of the initial and the final
neutral $B$ meson, one has -- in the absence of a
$B^0/\bar B^0$ production asymmetry -- to employ
flavour-specific decays of the beauty hadron that was
produced together with the $B^0$ or $\bar B^0$
meson; e.g. one measures
$B\, \bar B^0 \ra (l^+X)(l^+X)$ vs.
$\bar B\, B^0 \ra (l^-X)(l^-X)$.
However a very tiny difference in the two rates is
expected:
$$\frac{\Gamma (B\bar B^0\ra l^+l^+X)-
\Gamma (\bar BB^0\ra l^-X)}
{\Gamma (B\bar B^0\ra l^+l^+X)+
\Gamma (\bar BB^0\ra l^-X)}\sim 10^{-4}
\div 10^{-3}\eqno(14)$$

(vii) In the reaction $\Upsilon (4S)\ra B\bar B$ one
starts from a pure quantum state $J_{CP}=1^{--}$. One can still
define a CP asymmetry for the reaction rate yielding a charged
lepton at time $t_l$ and a final state like $\psi K_S$
at time $t_{\psi}$ \cite{CBS,BSKU}:
$$\Delta \Gamma (CP)\equiv \Gamma [\Upsilon (4S)
\ra (l^-X;t_l)_B(\psi K_S;t_{\psi})_B]-
\Gamma [\Upsilon (4S)
\ra (l^+X;t_l)_B(\psi K_S;t_{\psi})_B] $$
$$\propto \sin(\Delta m(t_l-t_{\psi}))
\Im \frac {q}{p}\bar \rho _{\psi K_S}\; .
\eqno(15)$$
The time {\it difference} $t_l-t_{\psi}$ enters because the
$B^0\bar B^0$ pair forms a C-odd state. Upon integrating over the
times of decay $t_l$ and $t_{\psi}$ such a CP asymmetry
necessarily has
to vanish:
$$\int dt_ldt_{\psi}\Delta \Gamma (CP)\equiv 0\; !\eqno(16)$$

(viii)  Production asymmetries, if properly
utilized, represent a great virtue rather than a vice.
The rate for the transitions of a beam of neutral $B$ mesons
into a CP eigenstate, like $\psi K_S$, as a function of the time of
decay is given by
$$\Gamma (B_{neutral}\ra \psi K_S;t)\propto e^{-\Gamma _Bt}
(1+P\cdot \Im \frac{q}{p}\bar \rho _{\psi K_S}\sin (\Delta mt))
\, , \eqno(17)$$
where
$$P\equiv \frac{N(B)-N(\bar B)}{N(B)+N(\bar B)}\eqno(18)$$
denotes the difference between the numbers of produced
$B^0$ and $\bar B^0$ mesons.
If $P\neq 0$, an observable CP asymmetry arises even without
independent flavour tagging! Furthermore the quantity $P$ can be
measured in other channels where no CP violation can occur:
$$\Gamma (B_{neutral}\ra \psi K^-\pi ^+;t)\propto
e^{-\Gamma _Bt}(1+P\cos (\Delta mt))\; , \eqno(19)$$
$$\Gamma (B_{neutral}\ra \psi K^+\pi ^-;t)\propto
e^{-\Gamma _Bt}(1-P\cos (\Delta mt))\; . \eqno(20)$$
Comparing $P$, as extracted from eqs. (19) and (20),
provides a check on possible detector biases -- as does,
to some extent,
$$\Gamma (B_{neutral}\ra \psi X;t)\propto
e^{-\Gamma _Bt}\, . \eqno(21)$$

\subsection{Parametric KM Predictions}

Up to now we have discussed how to measure CP asymmetries
in terms of $\Im (q/p)\bar \rho _f$. Next we will
analyse how well this quantity can be predicted in terms of
the KM parameters.

\vspace{.3cm}

(a) $B_d\ra \psi K_S$

\vspace{.3cm}

\noindent This mode, which is driven
by the quark level transition $b\ra c\bar cs$, is described
by a single isospin amplitude:
$$T(\bar B_d\ra \psi K_S)=V(cb)V^*(cs)e^{i\al _{1/2}}|M_{1/2}|
\; , \eqno(22a)$$
$$T(B_d\ra \psi K_S)=V^*(cb)V(cs)e^{i\al _{1/2}}|M_{1/2}|\, .
\eqno(22b)$$
Thus
$$\bar \rho (\psi K_S)=\frac {V(cb)}{V^*(cb)}\; \; \; \;
\Rightarrow \; \; \; \; |\bar \rho (\psi K_S)|=1
\, , \eqno(23)$$
i.e. no {\em direct} CP violation arises in
$B_d\ra \psi K_S$.

Since $q/p\simeq \sqrt {M^*_{12}/M_{12}}\simeq
V(td)/V^*(td)$, one arrives at \cite{CBS,BSKU}
$$\Im \frac {q}{p}\bar \rho (\psi K_S)\simeq
\Im \frac {V^2(cb)V^2(td)}{|V(cb)V(td)|^2}=\sin 2\phi _1
\eqno(24)$$
where the last equality follows from inspecting the KM triangle
of Fig. 1.

\vspace{.3cm}

(b) $B\ra \pi ^+\pi ^-$

\vspace{.3cm}

\noindent The quark level reaction
$b\ra u\bar ud$ underlying this transition contains
$\Delta I=1/2$ and $3/2$ contributions. The $\pi \pi$
final state thus carries isospin $0$ or $2$. Therefore
$$T(\bar B_d\ra \pi ^+\pi ^-)=e^{i\al _2}|M_2|V(ub)V^*(ud) +
e^{i\al _0}[|M_0|V(ub)V^*(ud)+\tilde M_0 V(cb)V^*(cd)]
\; , \eqno(25a)$$
$$T(B_d\ra \pi ^+\pi ^-)=e^{i\al _2}|M_2|V^*(ub)V(ud) +
e^{i\al _0}[|M_0|V^*(ub)V(ud)+\tilde M_0^* V^*(cb)V(cd)]
\; .\eqno(25b)$$
Here $\tilde M_0$ represents the `penguin' contribution and is
actually complex. Now one has $|\bar \rho (\pi \pi)|\neq 1$,
i.e. direct CP violation can arise in $B_d\ra \pi ^+\pi ^-$,
though it would be surprising if it amounted to a numerically
large effect. Accordingly one expects
$$\Im \frac{q}{p}\bar \rho (\pi ^+\pi ^-)\sim
\sin 2\phi _2\eqno(26)$$

(c) Similarly one finds (with the same caveats as for
$B_d\ra \pi ^+\pi ^-$)
$\Im (q/p)\bar \rho (B_s\ra K_S\rho _0)\sim \sin 2\phi _3$.

\vspace{.3cm}

The following global
judgement should be noted: the reliability with which one can
predict, in terms of the SM parameters,

\noindent -- the quantity $\Im (q/p)\bar \rho _f$
is between excellent and good,

\noindent -- the $B^0-\bar B_0$ oscillation rate
$(\Delta m_B)^{-1}$ is between good and decent, and

\noindent -- the relevant branching ratios is at best
decent.

\subsection{Numerical Predictions}

%{\em (1) Using present information on the KM parameters}

The baseline of the KM triangle in Fig. 1 can
conveniently be normalized
to unity without changing the angles and thus the CP
asymmetries. Determining two more elements will then
define the KM triangle. At present we have some
estimates on $|V(ub)/V(cb)|$ that suffer from considerable
experimental and theoretical uncertainties; some broad
bounds on $|V(td)|$ can be suggested and the value of
$\phi _1$ can be inferred. I will discuss here how much we
know about these quantities and how the situation will
presumably improve {\em before} the first significant
study of CP violation in $B$ decays. For proper perspective
I will first address the qualitative features.

The angle $\phi _1$ can be inferred from existing
measurements, albeit with large {\it theoretical}
uncertainties.
For large top quark masses $\epsilon _K$ -- like $\Delta m_B$ --
is dominated by the top quark contributions to
the quark box diagram, and one finds
$|\epsilon _K|/\Delta m(B_d)\propto \sin 2\phi _1$,
largely independent
of the value of $m_t$. More specifically one obtains
$$ \sin 2\phi _1\simeq 0.33\times UNC\; , \eqno(27a)$$
where the factor $UNC$ contains the numerical uncertainties:
$$UNC\simeq \left( \frac{0.045}{|V(cb)|}\right) ^2
\left(\frac{0.72}{x_d}\right)
\left(\frac{\eta ^{(B)}_{QCD}}{0.55} \right)
\left(\frac{0.62}{\eta ^{(K)}_{QCD}}\right)
\left( \frac{2B_B}{3B_K}\right)
\left(\frac{f_B}{160\, \MeV}\right) ^2\eqno(27b)$$
with $x_d\equiv \Delta M(B_d)/\Gamma _B$. The first two
factors in eq. (27b) represent the uncertainties in the measurements
of the $B$ lifetime and of $B_d-\bar B_d$ oscillations, the next two
factors the ones in the perturbative QCD corrections and the last two
those in the hadronic matrix elements. Obviously, the latter are
numerically the most significant ones. On the other hand,
the value of the top mass does $not$ enter in this
estimate directly. From eq. (27a) one obtains
$\phi _1\sim 10^{\circ}$; for $f_B=200$ MeV one actually gets
$\phi _1\simeq 16^{\circ}$ and $\sin 2\phi _1\simeq 0.52$!

The length of the
side BC of the normalized
triangle is determined by $|V(ub)/V(cb)|$.
To have (at least) one solution for the triangle one needs
$|V(ub)/V(cb)|\geq \lambda \sin \phi _1$. Due to the
uncertainties in the correct value of $\phi _1$ this is
not an overly useful constraint for the time being, but I will
come back to this point. Up to now one has relied largely on models
of uncertain reliability to extract $|V(ub)/V(cb)|$ from
the data. But the situation will improve systematically
by a judicious application of two complementary analyses,
namely by measuring
the ratio of the {\em exclusive} semileptonic
decay rates for $B\ra l\nu \rho$ and $B\ra l\nu D^*$ and using ideas
based on Heavy Quark Symmetry \cite{NEUBERT};
or from a detailed study of the
lepton spectrum in {\em inclusive} semileptonic $B$ decays employing
Heavy Quark Expansions \cite{PRL}.
For the latter analysis it would be highly
desirable to measure the photon spectrum in inclusive radiative
decays $B\ra \gamma +X_s$ to gain complete control over the
theoretical uncertainties \cite{MOTION}. Pursuing both lines of
analysis vigorously will enable us to reliably estimate the errors.
In any case, even with the most precise determination of
$|V(ub)/V(cb)|$ a two-fold ambiguity will
obviously remain in the triangle
thus constructed (except for the special case
$|V(ub)/V(cb)|=\lambda \sin \phi _1$
\footnote{We already know that $|V(ub)/V(cb)|<\lambda$.}):
in case I [case II] $\phi _2$ will be
larger [smaller]
than $90^{\circ}$, as illustrated in Fig. ~\ref{F3}.
I will return to
this distinction several times.

\begin{figure}
\begin{center}
\mbox{\epsfig{file=7207fg3.eps,height=4cm}}
%\vspace*{-1cm}
\end{center}
\caption[]{Case I and case II scenario for the KM triangle}
\label{F3}
\end{figure}

The length of the side AB can in principle be extracted from
the observed oscillation strength
$x_d=\Delta m(B_d)/\Gamma _B$, once the
top mass is known with some accuracy. Yet in practice its
extraction is beset by considerable uncertainties in the
proper values for $B_Bf_B^2$. However one can hope to be able
to decide between the two cases I and II listed above,
obtained from a rather precise extraction of $|V(ub)/V(cb)|$.
A measurement of the branching ratio for the very rare
$K$ decay mode $K^+\ra \pi ^+ \nu \bar \nu$ -- expected to
be of order a few$\times 10^{-10}$ -- would enable us
to determine $|V(td)|$ very reliably \cite{GABBIANI}.

Measuring the speed of $B_s-\bar B_s$ oscillations
would also provide us with information on
$|V(td)|$ \cite{ALI}:
$$\frac{x(B_s)}{x(B_d)}\simeq \frac{\Delta m(B_s)}
{\Delta m(B_d)}=\frac{Bf_B^2(B_s)}{Bf_B^2(B_d)}
\cdot \frac{|V(ts)|^2}{|V(td)|^2}\geq
\frac{|V(ts)|^2}{|V(td)|^2}\simeq
\frac{|V(cb)|^2}{|V(td)|^2}\eqno(28)$$

\vspace{.3cm}
Three elements of the
relevant KM triangle are thus known to some degree:

\noindent (i) the baseline, since it can be normalized to unity;

\noindent (ii) the angle $\phi _1$, which can be inferred from the
data on $\epsilon _K/\Delta m(B_d)$;

\noindent (iii) the side $BC$, which depends on $|V(ub)/V(cb)|$.

Item (ii) is beset by large theoretical uncertainties,
whereas item (iii) at present suffers from quite
considerable experimental uncertainties as well.
The latter situation can be expected to improve significantly
in the foreseeable future. The KM triangle will then be determined
up to a two-fold ambiguity: a case I with the corners
$AB^IC$ and a case II with $AB^{II}C$, Fig. 3. Due to
$|V^I(td)| < |V^{II}(td)|$ case I
requires higher values for $m_t$ and/or $f_B$ to reproduce the
observed magnitude of $x(B_d)$ than case II. By the same token
case I leads to faster $B_s-\bar B_s$ oscillations
than case II, and $x^I_s>(f(B_s)/f(B_d))^2\cdot \lambda ^2
\cdot \cos ^2\phi _1 > 20\cos ^2\phi _1$. Lastly,
$\phi _2^I<90^{\circ}$ versus
$\phi _2^{II}>90^{\circ}$ and thus
$\sin 2\phi _2^I>0$ in contrast with $\sin 2\phi _2^{II}<0$; i.e.
the CP asymmetry predicted for $B_d\ra \pi ^+\pi ^-$
differs {\em in sign} for the two cases!

These general considerations help us understand
more directly the results of a numerical
phenomenological analysis:
in Fig. ~\ref{F4A} the allowed forms of the KM triangle
are shown for $140\, \MeV \leq f_B\leq 240\, \MeV$,
$0.05\leq |V(ub)/V(cb)|\leq 0.11$ and
$130\, \GeV\leq m_{top}\leq 190\, \GeV$. The top of the triangle
has to lie in the shaded area. A wide variation in the values for
the three angles $\phi _i$ is then allowed.
Figure ~\ref{F4B}
illustrates the future situation when $m_{top}$ and
$|V(ub)/V(cb)|$ have been measured to, say, 10\% accuracy,
e.g. $m_t=160\pm 10\, \GeV$ and
$|V(ub)/V(cb)|=0.08\pm 0.01$. A significant improvement
emerges; yet one should note that
there are two disjoint areas where
the top of the triangle is allowed to lie. The two-fold ambiguity
mentioned above thus arises even when $m_t$ and
$|V(ub)/V(cb)|$ are fairly well measured -- unless one can
decide between $f_B > 210\, \MeV$ and $f_B< 170\, \MeV$.

\begin{figure}
\begin{center}
\mbox{\epsfig{file=7207fg4.eps,height=6cm}}
%\vspace*{-1cm}
\end{center}
\caption[]{Shape of the KM triangle inferred from present
phenomenology}
\label{F4A}
\end{figure}

\begin{figure}
\begin{center}
\mbox{\epsfig{file=7207fg5.eps,height=6cm}}
%\vspace*{-1cm}
\end{center}
\caption[]{Shape of the KM triangle after future measurements
of V(ub)/V(cb) and $m_t$}
\label{F4B}
\end{figure}

There is one test to which the trigonometry of the KM
triangle could be subjected {\em before} the first
measurement of a CP asymmetry in $B$ decays: once
a value for $|V(td)|$ is extracted from
$\Delta m(B_s)/\Delta m(B_d)$ or from
$BR(K^+\ra \pi ^+\nu \bar \nu)$, the triangle is defined
\footnote{ I assume here that $|V(td)/(\lambda V(ts))|
\geq 1- |V(ub)/(\lambda V(cb))|$ holds, since
no triangle can be constructed otherwise; in that case the
incompleteness of the Standard Model would have been
established.}.
With $m_{top}$ known, one can then deduce the value for
$Bf_B$ which is necessary to reproduce $\Delta m(B_d)$.
This, in turn, will allow us to infer quite reliably
$\sin 2\phi _1$ from $\epsilon _K/\Delta m(B_d)$, as
given above, and compare it with the value obtained from the
KM triangle!

\subsection{Tests of the KM Triangle}

{\em (i) The first significant CP study in $B_d$ decays}

\vspace{.3cm}

It appears quite likely that the first significant study of
CP violation involving $B^0-\bar B^0$ oscillations will
be performed in $B\ra \psi K_S$ decays. If it is found that
the value for $\sin 2\phi _1$ {\it measured} differs from the one
{\it inferred} from $\epsilon _K/\Delta m(B_d)$, i.e.
$$\sin 2\phi _1|_{measured}\; \; \neq
\sin 2\phi _1|_{inferred}\, , \eqno(29)$$
it would be tempting to invoke the intervention of `New
Physics' as explanation. Yet this would be justified only if
$|V(td)|$ had previously been determined from $x_s/x_d$ or
from $BR(K^+\ra \pi ^+\nu \bar \nu)$ and incorporated into
$\sin 2\phi _1|_{inferred}$; or if the bound
$|V(ub)/V(cb)|\geq \lambda \sin \phi _1$
were found to be violated.
Otherwise one would have to adopt the
more conservative -- and
therefore more appropriate -- interpretation
that {\em incorrect} values for $Bf_B^2/B_K$ had been
used in deducing $\sin 2\phi _1$ from
$\epsilon _K/\Delta m(B_d)$. On the other hand,
with $\phi _1$ thus measured, an important milestone
will have been achieved: the {\em normalized} KM triangle will
then have been defined completely through measurements of
its elements in the $B$ system, namely $|V(ub)/V(cb)|$ and
$\sin 2\phi _1$. The `umbilical cord' to CP violation in
$K$ decays will
have been severed and $B$ physics will have achieved
`autonomy'! A first indirect self-consistency check
will arise as a by-product: from the measured value of
$\sin 2\phi _1$ one can deduce the required value of
$Bf_B^2$, using eqs. (27); with $m_t$ known at that time,
as expected, one can then check whether the value for
$|V(td)|$ extracted from $\Delta m(B_d)$ is consistent
with the KM triangle in either case I or case II.

\vspace{.3cm}

{\em (ii) Search for New Physics}

\vspace{.3cm}

If the {\em measured} CP asymmetry in $B\ra \pi ^+ \pi ^-$
turns out to be different from the value obtained
from the KM triangle, i.e. $if$
$$ \sin 2\phi _2|_{B_d\ra \pi \pi}\neq
\sin 2\phi _2|_{KM \; \Delta}\, , \eqno(30)$$
then one would have established the intervention
of New Physics. In this context one should keep in
mind that such a CP asymmetry, which depends on the
{\it ratio} of amplitudes, possesses a
sensitivity to New Physics considerably higher
than other indirect probes such as
branching ratios that depend on the squares of amplitudes.
Later on I will cite some examples.

The hypothesis that the KM ansatz provides the only source
of CP violation in $B$ decays is tested most cleanly by
checking whether the KM triangle closes properly:
$$\phi _1 + \phi _2 + \phi _3 = 180 ^{\circ}
\, . \eqno(31)$$
In principle one can determine $\sin 2\phi _3$ from
the CP asymmetry in the KM-suppressed
mode $B_s\ra K_S\rho ^0$. In practice,
however, this amounts to an even higher challenge
than extracting $\phi _2$ from $B\ra \pi \pi $ decays:
the production cross section is considerably lower and
the $B^0-\bar B^0$ oscillation rate, which threatens to
wash out the asymmetry, is expected to be much higher in
$B_s$ than in $B_d$ decays.

At this point one has to keep the following
in mind:  within the SM
$\phi _3$ is of course
known to be $180^{\circ}-\phi _1-\phi _2$,
and it is quite unlikely that its value
could be extracted from $B_s\ra K_S\rho ^0$
with an accuracy comparable to that in $\phi _1$
and $\phi _2$. The motivation behind measuring
$\phi _3$ independently is to search
for the intervention of New Physics. It then makes
more sense to analyse a transition that commands a
higher branching ratio and a cleaner experimental
signature as well as theoretical interpretation.
The KM-allowed modes $B_s\ra \psi \phi,\, \psi \eta$
fit this bill quite nicely: their branching ratio is
expected to be around 0.2\%; the $\psi $ decay
provides a striking signature; finally a small CP
asymmetry, say below 2\%, is predicted by the
specific features of the KM ansatz. The last point is
easily understood \cite{BS}:
for in the transitions $B_s\ra \psi \phi$
and
$B_s\Rightarrow \bar B_s \ra \psi \phi$ quarks of only
the second and third family contribute.
Thus there can be no observable CP violation,
apart from violations of weak universality. CP
asymmetries are then suppressed by $\lambda ^2$.
However once New Physics intervenes in the
$\Delta B=2$ sector driving $B^0-\bar B^0$ oscillations,
it will in general generate a CP asymmetry that has, in
$B_s\ra \psi \phi$ decays, a nearly zero SM
background -- in contrast with $B_d\ra \psi K_S$:
$$\Im \frac{q}{p}\bar \rho (B_s\ra \psi \phi)=
\sin 2\zeta ^{NP}\, . \eqno(32)$$

\section{New Physics Scenarios}

It should be recalled that the lessons one had learnt from
a detailed study of the dynamics of strangeness turned out to
be crucial in formulating the Standard Model: analysis of
$\Delta S=1,0$ processes lead to the concepts of
flavour quantum numbers and Cabibbo universality
and it suggested the occurrence of parity violation;
a study of $\Delta S=2$ transitions lead to postulating
the existence of charm quarks (based on the absence of
flavour-changing neutral currents) and of top quarks (due to CP
violation).

There is every reason to expect that history will repeat
itself, in the sense that a comprehensive study of beauty
transitions will reveal the existence of a `New Standard
Model'. It is conceivable -- though not very likely --
that NP will manifest itself in tree-level $\Delta B=1$
decays; for instance, it has been suggested that the $b$ quark
possesses {\em right-}handed charged current couplings
\cite{GRONAU}.
The prospects are much brighter for NP revealing
itself in $\Delta B=2$ transitions: these processes are highly
forbidden in the SM and thus exhibit a large sensitivity to
high-mass virtual states. NP
can change the relation between $\Delta m(B_d)$ and
$\Delta m(B_s)$ and between the different CP asymmetries
in $B_d$ and $B_s$ decays. Since we are dealing here with
coherent processes, these effects depend on the NP amplitude
directly, and not only on its modulus squared.

Such a program will be illustrated here with the
example of SUSY. For our purposes the most
relevant point of SUSY is that it induces flavour-changing
neutral currents in the strong couplings of gluinos to
quarks and squarks \cite{SUSY}.
The quantity $M_{12}(B^0)$ then
receives contributions of order $\al _S^2$ from SUSY in
contrast to order $\al _{weak}^2$ terms from SM. Those
NP contributions can then be numerically significant,
even for gluino and squark masses as large as $300 -- 400$
GeV.

In minimal SUSY, which is the `Standard Extension' of the
Standard Model, the flavour-changing squark couplings
are controlled by the corresponding KM parameters,
and the CP asymmetries are affected only {\em indirectly}.
Yet once non-minimal SUSY is considered,
a veritable Pandora's box for
new weak phases opens up \cite{GABBIANI2}. One finds
$$\Im \frac{q}{p}\bar \rho (B_d\ra \psi K_S)\equiv
\sin (2"\phi _1")=
\sin (2\phi _1|_{KM}+\xi _{SUSY}(B_d))\eqno(33)$$
$$\Im \frac{q}{p}\bar \rho (B_d\ra \pi ^+ \pi ^-)
\equiv
\sin (2"\phi _2")=
\sin (2\phi _2|_{KM}-\xi _{SUSY}(B_d))\eqno(34)$$
$$"\phi _1" +"\phi _2"=\phi _1+\phi _2\, , \eqno(35)$$
where $\xi _{SUSY}(B_d)$ denotes the phase generated
by the gluino-squark box diagram contribution to
$M_{12}(B_d)$. It therefore affects $B\ra \psi K_S$
and $B\ra \pi ^+\pi ^-$ decays in the same way, which
is expressed by eq. (35). There exist then two ways
to search for a `smoking gun':

\noindent (i) With "$\phi _1$" and "$\phi _2$" measured,
the normalized KM triangle is determined. Experimental
information on $|V(ub)/V(cb)|$ and/or $m_{top}$ will
then provide a check on the KM trigonometry.
Any resulting inconsistency reveals the intervention of NP.

\noindent (ii) For the CP asymmetry in $B_s\ra K_S\rho ^0$
one obtains:
$$\Im \frac{q}{p}\bar \rho (B_s\ra K_S\rho ^0)\equiv
\sin 2"\phi _3"=\sin 2(\phi _3|_{KM}+\xi _{SUSY}(B_s))
\eqno(36)$$
with in general $\xi _{SUSY}(B_d)\neq \xi _{SUSY}(B_s)$.
Altogether one finds:
$$"\phi _1"+"\phi _2"+"\phi _3"=
\phi _1|_{KM}+\phi _2|_{KM}+\phi _3|_{KM}+
\xi _{SUSY}(B_s)=180^{\circ}+\xi _{SUSY}(B_s)
\neq 180^{\circ}\, !
\eqno(37)$$
As already mentioned, a much more promising way to search
for a `smoking gun' is to analyse $B_s\ra \psi \phi,\,
\psi \eta$ decays, since
$$\Im \frac{q}{p}\bar \rho (B_s\ra \psi \phi )\simeq
\sin 2\xi _{SUSY}(B_s)\, . \eqno(38)$$

\section{Summary on $B^0$ Decays}

The salient features of the preceding sections are:

\vspace{.3cm}

\noindent $\bullet$ The SM unequivocally predicts large CP
asymmetries in certain exclusive non-leptonic $B$ decays of order
several$\times$10\%, i.e. two orders of magnitude larger
than in $K\ra \pi \pi$.

\noindent $\bullet$ Those huge CP asymmetries involve
$B^0-\bar B^0$ oscillations and possess a very peculiar
dependence on the (proper) time of decay:
$$\Delta \Gamma (CP)\propto e^{-\Gamma _Bt}\cdot
\sin \Delta m_Bt \cdot
\Im \frac{q}{p}\bar \rho _f\; . \eqno(39)$$

\noindent $\bullet$ They occur in two distinct classes
of $B_d$ decays, driven by different quark level transitions:
$$\Im \frac {q}{p}\bar \rho _f(b\ra c\bar cs)=
\sin 2\phi _1\eqno(40)$$
$$\Im \frac {q}{p}\bar \rho _f(b\ra u\bar ud)\simeq
\sin 2\phi _2\; .\eqno(41)$$
In general one has
$$\sin 2\phi _1\neq \sin 2\phi _2\, , \eqno(42)$$
i.e. there is large {\em direct} CP violation; the KM ansatz
does {\it not} provide a superweak scenario for $B$ decays!

\noindent $\bullet$ In each decay class there are several
useful modes where a CP asymmetry can occur:
$$b\ra c\bar cs:\; \; \; \; B_d\ra \psi K_S, \psi K_S\pi ^0,
\psi \pi \pi ,D\bar D,...\eqno(43)$$
$$b\ra u\bar ud:\; \; \; \; B_d\ra \pi ^+\pi ^-,
\pi ^0\pi ^0,\rho ^0\rho ^0, a_1\pi, ...\eqno(44)$$

\noindent $\bullet$ The CP asymmetry for $b\ra c\bar cs$ type
decays can be expressed reliably in terms of the KM parameters
without any uncertainty from hadronic matrix elements.
For $b\ra u\bar ud$ type decays on the other hand, complications
due to FSI can in principle arise; yet those aspects can be
probed experimentally. The first warning sign goes up if one
finds significant {\em direct} CP violation in
$B\ra \pi ^+\pi ^-$ decays.

\noindent $\bullet$ Such CP studies are -- with presently
available technology for vertex detection --
not feasible at a {\em symmetric}
$\Upsilon (4S)\ra B\bar B$ factory.

\noindent $\bullet$ There exist valuable cross-checks against
detector biases: e.g. {\em no genuine} differences
can exist in
$$B_d\ra \psi +X\; \; \; \; vs. \; \; \; \;
\bar B_d\ra \psi $$
or in
$$B_d\ra \psi K^-\pi ^+\; \; \; \; vs. \; \; \; \;
\bar B_d\ra \psi K^+\pi ^-\eqno(45)$$

\noindent $\bullet$ There exists high sensitivity to NP
`lurking' in $\Delta B=2$ dynamics; it can reveal itself
through

\noindent -- inconsistencies in KM trigonometry or

\noindent -- observable CP asymmetries in
$B_s\ra \psi \phi,\, \psi \eta$ decays.

\noindent $\bullet$ Once the results of
such a CP analysis are
taken together with
quantitative data on $BR(b\ra s\gamma )$,
$BR(b\ra sl^+l^-)$, $m_{top}$ and $\Delta m(B_s)$,
one would probably be able to identify some definite
clues about the type of NP that is intervening:
SUSY, horizontal interactions, a
non-minimal Higgs sector, etc.

\section{Direct CP Violation}

There are various significant features that make searches
for direct CP violation easier than the studies outlined
above:
the decay modes are flavour-specific
and thus self-tagging; since the asymmetry is independent of the
time of decay, there arise considerably lower demands on
tracking the $B$ decay vertices, and meaningful searches
can be performed at a symmetric $\Upsilon (4S)$ $B$ factory;
the asymmetries depend directly on $\phi _3$. There are, of
course, serious drawbacks as well: the relevant modes are
necessarily KM-suppressed; the asymmetries are not sensitive
to $\phi _1$ or $\phi _2$, and hardly to NP; they require --
for rate asymmetries -- the intervention of non-trivial
FSI, which tends to put them beyond theoretical control;
the signal, which is just a rate difference independent of the
time of decay, is much more easily faked by background.

\begin{figure}
\begin{center}
\mbox{\epsfig{file=7207fg6.eps,height=5cm}}
\end{center}
\caption[]{Quark diagrams for $B^-\ra D^0K^-$, (a),
and $B^-\ra \bar D^0K^-$, (b)}
\label{F5}
\end{figure}

I will sketch here only one example, which
is the cleanest one from a theoretical
perspective, namely how to extract $\phi _3$
from a comparison of $B^-\ra D^0/\bar D^0K^-$ and
$B^+\ra D^0/\bar D^0K^+$ \cite{PAIS}.
In Fig. ~\ref{F5} the quark diagrams for
$B^-\ra D^0K^-$
and for $B^-\ra \bar D^0K^-$, respectively, are shown.
They depend on different weak parameters, the former on
$V(cb)V^*(us)$, the latter on $V(ub)V^*(cs)$, and
their difference in the weak phase is given by $\phi _3$.
Furthermore $|V(cb)V(us)|\sim {\cal O}(\lambda ^3)\sim
|V(ub)V(cs)|$; thus the two amplitudes are -- except for the
colour factors -- comparable in magnitude, which enhances the
weight of their interference, if it occurs. The isospin of the two
final states is different, and one can expect non-trivial FSI to be
present. Thus all the ingredients are there for the existence of
a direct CP asymmetry -- with, at first sight, one crucial
exception: the $D^0$ and the $\bar D^0$ are different mesons,
which eliminates any interference between the two amplitudes!
Yet quantum mechanically the situation is actually more
subtle: the identity of the neutral charm meson as a
$D^0$ rather than a $\bar D^0$ can reveal itself
only through a flavour-specific decay, like
$D^0\ra K^-\pi ^+,\, K^{-*}\pi ^+$
\footnote{Strictly speaking, the final states
$K^-\pi ^+$ etc. provides `only' a 99\% efficient
$D^0$ filter because of doubly Cabibbo suppressed
$D$ decays; yet this is obviously
sufficient for practical purposes.}.
There is a subclass
of decays which as a matter of principle does not allow this
distinction (in the absence of sizeable CP asymmetries in
$D^0$ decays):
$$D^0\ra K_S\pi ^0,K_S\eta , K_S\rho ^0,K_S\omega ,
K_S\phi , ..., K\bar K,\pi \pi , ... \leftarrow
\bar D^0\; . \eqno(46)$$
It should be noted in passing that this is
the same quantum mechanical concept of
particle--antiparticle identity that opens the door for
$B^0-\bar B^0$ oscillations and, applied to the
$K^0-\bar K^0$ system, represents a pre-condition
for the observability of a CP asymmetry in
$B_d\ra \psi K_S$!
Accordingly we define CP eigenstates of neutral $D$ mesons:
$$|D_{\pm}\rangle =\frac{1}{\sqrt{2}}(|D^0\rangle
\pm |\bar D^0\rangle ) \eqno(47)$$
with
$$D_+\ra K\bar K,\, \pi \pi\; \; \; ,\; \; \;
D_-\ra K_S\pi ^0,\, K_S\eta ,\, K_S\omega ,\,
K_S\phi\; , \eqno(48)$$
and one derives easily
$$\Gamma (B^+\ra D_{\pm}K^+)-
\Gamma (B^-\ra D_{\pm}K^-)\simeq 2\sin \phi _3\cdot
\sin \Delta \al \cdot M_1\cdot M_2\; . \eqno(49)$$
There are four unknowns in eq. (49): the sought-after
KM angle $\phi _3$; the phase shift $\Delta \al$ between
the $D^0K^-$ and $\bar D^0K^-$ final states, and the two
reduced real matrix elements $M_1$ and $M_2$. Fortunately
there are four independent observables, namely
\cite{WYLER}
$$\Gamma (B^-\ra D_{\pm}K^-)\eqno(50a)$$
$$\Gamma (B^+\ra D_{\pm}K^+)\eqno(50b)$$
$$\Gamma (B^-\ra D^0K^-)=
\Gamma (B^+\ra \bar D^0K^+)\eqno(50c)$$
$$\Gamma (B^-\ra \bar D^0K^-)=
\Gamma (B^+\ra D^0K^+)\; ,\eqno(50d)$$
where $D^0$ and $\bar D^0$ are identified by their
flavour-specific decay modes. It is CPT invariance
that enforces the equality of the CP conjugate rates
in eqs. (50c,d).

The procedure to follow is in principle straightforward:
since $\Gamma (B^-\ra D^0K^-)=M_1^2$ and
$\Gamma (B^-\ra \bar D^0K^-)=M_2^2$, one can extract
$M_1$ and $M_2$ directly from these two decays.
Then one solves for $\phi _3$ and $\Delta \al$ from
eqs.(50a,b).

Extracting $\phi _3$ from this asymmetry is thus clean;
the price to be paid is that one has to deal with a small
effective branching ratio of $\sim 10^{-5}$ or less.

It is at least amusing to note that for the special value
$\phi _3=90^{\circ}$ the direct CP asymmetry is maximal in the
sense of $\sin \phi _3=1$, while it vanishes for a CP
asymmetry involving $B^0-\bar B^0$ oscillations --
$\sin 2\phi _3=0$!

\section{Summary and Outlook}

It is not my intent to deny that detailed and meaningful
searches for CP violation in beauty decays are extremely
challenging experimentally; instead I want to emphasize that
they are worth every effort! For the $B$ system is (almost)
optimal for CP studies: the lifetimes are long by the standards
of today's technology; $B^0-\bar B^0$ oscillations proceed in a
speedy fashion, the relevant weak phases are naturally large,
and at least some of the KM predictions for large CP asymmetries
are given with high parametric reliability. One must understand
that no other system with similarly large effects will come
along, certainly not the top system and also not the charm
sector.

Furthermore, the insights to be gained here are of fundamental
importance; they cannot be obtained any other way nor can they
become obsolete by any other development or discovery.
For instance,
while the discovery of the SM Higgs will have no impact
on the CP phenomenology, the measurement of the top mass
will sharpen some of the numerical predictions, in particular
concerning $\sin 2\phi _2$ and $\sin 2\phi _3$; a discovery of
charged Higgs and/or SUSY fields will provide crucial input for
interpreting the NP CP phenomenology.

Maybe the clearest evaluation of the importance of CP studies
in beauty decays can be given by interpreting the possible results of
such an enterprise. There are three possible outcomes:

(A) {\em No CP asymmetries are observed in $B$ decays!} To be
more specific, let us assume that an upper bound of 4\%
has been placed on all of them. Considering that there
are three families, that $B^0-\bar B^0$ oscillations are speedy
and based on what we already know about the KM parameters,
I regard this scenario as the least likely outcome. Yet
in that case we would have established that the KM ansatz is
{\em not} behind $K_L\ra \pi \pi$, that it is not even a
significant factor there! Negative results, with the exception
of the Michelson--Morley experiment, do not bestow glory upon
the responsible researchers; nevertheless in this case it
would constitute an important discovery: it would force us to
abandon the Standard Model paradigm of CP violation
and to attribute
the {\em observed} CP violation in $K_L$ decays to as yet
unidentified New Physics, with two immmediate
consequences: (i) It would lead to the formulation of a baffling
theoretical puzzle, namely `Why is there no significant KM
source for CP violation?' (ii) On the experimental side it would
lend new urgency to the need to search for different
manifestations of CP violation in light-quark systems,
such as the
electric dipole moments of neutrons or electrons,
the transverse polarization
of muons in $K\ra \mu \nu \pi$ decays,
$BR(K_L\ra \pi ^0l^+l^-)$, etc.

(B) {\em Large CP asymmetries are indeed found -- but they are
not (quite) consistent with the KM predictions!} This scenario --
which I regard as the most likely outcome -- will represent
indirect, yet unequivocal evidence for the
intervention of NP in $B$ transitions. Taken together with
measurements of $m_{top}$, $\Delta m(B_s)$, $BR(b\ra s\gamma )$,
$BR(b\ra sl^+l^-)$ (and hopefully
$BR(K^+\ra \pi ^+\nu \bar \nu )$ and $BR(b\ra d\gamma )$) it would
provide us even with some definite clues about the nature of the
NP involved. It would also allow us to check to which degree the
KM ansatz contributes to $K_L\ra \pi \pi$.

(C) {\em Large CP asymmetries in universal quantitative agreement
with the KM predictions are found, without any significant evidence
for NP -- even when extrapolating down to $K_L$ decays!} This
would mean much more than yet another success for SM like
others before. For one thing, it would confirm an `honest'
prediction -- rather than a postdiction -- of the KM mechanism
concerning a completely different dynamical system. Secondly --
and even more importantly -- it would provide the seeds for more
profound knowledge, since it addresses one of the most
central mysteries of the Standard Model, namely
the generation of fermion masses.
I will explain this point in more detail now.
After a comprehensive analysis of $B$ decays the\ KM matrix will be
fully known. Keep in mind that a non-trivial KM matrix arises
because the mass matrices for the three up-type and the three
down-type quarks are diagonalized by two different unitary matrices,
$R_{up}$ and $R_{down}$, respectively: $V_{KM}=R^*_{up}R_{down}$.
Thus $V_{KM}$ contains unique information on the quark mass matrices
{\it over and above} the quark mass
eigenvalues. To understand this
connection better, let us count the number of independent
parameters: The two $3\times 3$ quark mass matrices contain
13 {\it physical} parameters; on the other hand, there are 10
observable parameters, namely 6 quark masses and 4
KM parameters.
That means that in general not all elements of the quark mass
matrices can be determined through observation. However based
on some symmetry considerations one can postulate special values
for at least some of the elements of the quark mass matrices. As
it happens, theorists are not always very imaginative, and the
best they have come up with so far, is to require certain elements
to vanish. One can have at most six such `texture zeroes';
there can thus be as few as seven independent parameters in the
quark mass matrices, and then there arise relations between the
quark masses and the KM parameters:
$(V_{KM})_{ij}=f(M_{up}, M_{down})$. In the last step one evolves
these relations from the GUT scale, where they are postulated,
down to the hadronic scales probed in heavy flavour decays.
A comprehensive study of this kind has been given in
ref. \cite{RAMOND}; I summarize their results
for $m_{top}=180\, \GeV$ in a way that is convenient
for my discussion:

\vspace{0.5cm}

\begin{center}
\begin{tabular}{|l|l|l|l|l|l|} \hline
                         &A     &B    &C    &D    & E    \\
\hline
$|V(ub)|/|V(cb)|$        &0.06  &0.062&0.068&0.059&0.089 \\
\hline
$x_s/x_d$                & 23   &23   &21   &28   &34    \\
\hline
$\sin 2\phi _1$          & 0.52 &0.54 &0.58 &0.51 &0.72  \\
\hline
$\sin 2\phi _2$          &--0.17 &--0.22&0.31 &--0.39&--0.61 \\
\hline
$\sin 2\phi _3$          & 0.66 &0.71 &0.31 &0.81 &0.99  \\
\hline
$\sin  \phi _3$          & 0.94 &0.92 &0.99 &0.89 &0.75  \\
\hline
\end{tabular}
\end{center}

\vspace{0.5cm}

\noindent
where I have asssumed $(f(B_s)/f(B_d))^2 =1.1$ to translate
$|V(ts)/V(td)|^2$ into $x_s/x_d$. The letters A -- E refer
to five classes of mass matrices with texture zeroes in different
positions. The details are not important here, and I anticipate
considerable theoretical evolution and hopefully
progress in this field over the next five
years or so; but one can use these numbers to illustrate
important benchmarks. Scenario E is most easily distinguished
against the others by measuring $|V(ub)/V(cb)|$ and
$\sin 2\phi _1$; scenarios A--D on the other hand
basically agree on these two quantities, with
$\sin 2\phi _1$ being uniformly large. Considerable
differences arise in the values for $\sin 2\phi _2$,
even the sign can change! Scenarios A and B, which
are the closest numerically, differ there by 20\%.
I take these examples to indicate the
general trend and conclude:

\noindent $\bullet$ Both $\sin 2\phi _1$ {\em and}
$\sin 2\phi _2$ have to be measured -- that is
non-negotiable!

\noindent $\bullet$ While one aims for a 20--30\% accuracy
for the first and second round measurements, the ultimate
goal is to achieve a 5\% precision or better to exploit the
discovery potential to the fullest.

Let me give an overview of how I anticipate the field to
proceed through three stages:

(A) `The era of explorers, trappers and traders': This is the
period where LEP, CDF, D0, HERAB and CLEO III will have a first
go at measuring $\Delta m(B_s)$ and some CP asymmetries. Like
in the days of the Old West in North America it is a period
where some expeditions that start out on their journey are never
heard of again. But those that do return with their spoils become
part of the legends.

(B) `The era of railroads and railroad barons': This is the period
where companies and people -- like Sen. Stanford who later
founded Stanford University -- with some capital set
out to lay down railroad tracks across the vast
expanses of land, guided by their vision, not afraid of risks
and not overly concerned about the survival of buffaloes. It will
be the SLAC and KEK asymmetric $B$ factories that will map out
the topography of the KM land.

(C) `The era of Multinational Corporations': It is then that
the vast new territories, opened up by the railroads,
will be exploited to the fullest, i.e. when LHC will
hopefully allow the ultimate measurements to be made.

I would like to emphasize that the
usefulness of era (A) or era (B)
is not removed by the promises of the
subsequent era; likewise
the importance of era (B) or era (C)
is {\em not} conditioned by the
failure of the preceding one. To cite but one example:
while no systematic study has been made yet, it seems to me
that the irreducible theoretical uncertainties will
ultimately arise on the
per cent level. Thus experimental studies have
to reach the 1\% level in precision before one has
exhausted the vast discovery potential in beauty
decays!

There is clearly a very long campaign ahead of us.
Let us keep in mind some lessons from Alexander the
Great's success: it is the training and motivation
of the army that is decisive, rather than its sheer size;
and you have to have the courage to set out on your long march!

\vspace*{0.5cm}

{\bf Acknowledgements:} \hspace{.4cm}
Mario Greco and Giorgio Bellettini have once again succeeded in
creating a very stimulating meeting in a pleasant setting, which
I have enjoyed tremendously. I am grateful to Ian Scott
for help with the figures.
This work was supported by the National Science Foundation under
grant number PHY 92-13313.

\vspace*{3cm}
\end{document}